\documentclass[aps,nofootinbib]{revtex4}
\usepackage[english]{babel}
\usepackage{color}
\usepackage{graphicx}
\usepackage{amsmath}
\usepackage{subcaption}
\begin{document}
\title{Simple analytic equation for airfoil shape description}
\author{David Ziemkiewicz}
\email{david.ziemkiewicz@utp.edu.pl}\affiliation{ Institute of Mathematics and
Physics, UTP University of Science and Technology, Al. Prof. S.
Kaliskiego 7, PL 85-789 Bydgoszcz
 (Poland)}
\begin{abstract}  
We show a simple, analytic equation describing a class of two-dimensional shapes well suited for representation of aircraft airfoil profiles. Our goal was to create a description characterized by a small number of parameters with easily understandable meaning, providing a tool to alter the shape with optimization procedures as well as manual tweaks by the designer. The generated shapes are well suited for numerical analysis with 2D flow solving software such as XFOIL.\end{abstract} \maketitle
\section{Introduction}
A numerically fast method of obtaining aerodynamic characteristics of an aircraft wing airfoil, e. g. the shape of the wing as seen in cross - section, is of great interest, especially since obtaining an exact solution based on full Navier - Stokes equations is usually a prohibitively complex task. Over years, a range of approaches has been developed to predict the lift and drag forces acting on a given shape. The analytical solutions are possible with the conformal mapping, where a solved problem of a flow around a simple shape such as a cylinder is used along coordinate transform to obtain the lift and drag of an airfoil shape \cite{Benson}. Nowadays, the analysis of a two-dimensional viscous and inviscid flow over an airfoil can be performed with a great accuracy on a personal computer, and the so-called inverse design can be used to generate the shape out of a prescribed pressure distribution \cite{Drela}. Alternatively, one can design the airfoil by trial and error, using some model equation to generate the geometry and refine it either manually or with optimization procedure. Therefore, it is of interest to provide a model having enough flexibility to describe a wide range of airfoils while keeping the number of free parameters to the minimum. Several commonly used approaches for shape generation and genetic optimization \cite{Goldberg} are B-splines \cite{Fanjoy,Viccini} and other parametric descriptions such as 11 parameter PARSEC airfoil family \cite{Sobieczky}. 

\section{Airfoil Equation}
The airfoil is defined as a pair of parametric equations for $X(\theta)$ and $Y(\theta)$ for $\theta=\langle 0,2\pi \rangle$,
\begin{eqnarray}\label{Main_eq}
X(\theta)&=&0.5 + 0.5\frac{|\cos \theta|^{B}}{\cos \theta}, \\
Y(\theta)&=&\frac{T}{2}\frac{|\sin \theta|^{B}}{\sin \theta}(1-X^P)+C\sin(X^E\pi)+R\sin(X2\pi),
\end{eqnarray}
where\\\\
B - Base shape coefficient. For B=2, the base shape of the airfoil is ellipse. In the limit of B approaching 1, the shape becomes a rectangle. This parameter affects mostly the leading edge.\\\\
T - thickness as a fraction of the chord.\\\\
P - taper exponent. For P=1, when going to the trailing edge, the thickness approaches 0 in a linear manner. For higher value, the airfoil tapers more suddenly near the trailing edge.\\\\
C - camber, as a fraction of chord.\\\\
E - camber exponent, defining the location of the maximum camber point, where $E=1$ describes $50\%$ camber point, and smaller value shifts it towards the leading edge.\\\\
R - reflex parameter. Positive value generates reflexed trailing edge, while negative one emulates flaps.\\\\

\begin{figure}
\centering
\includegraphics[width=.15\linewidth]{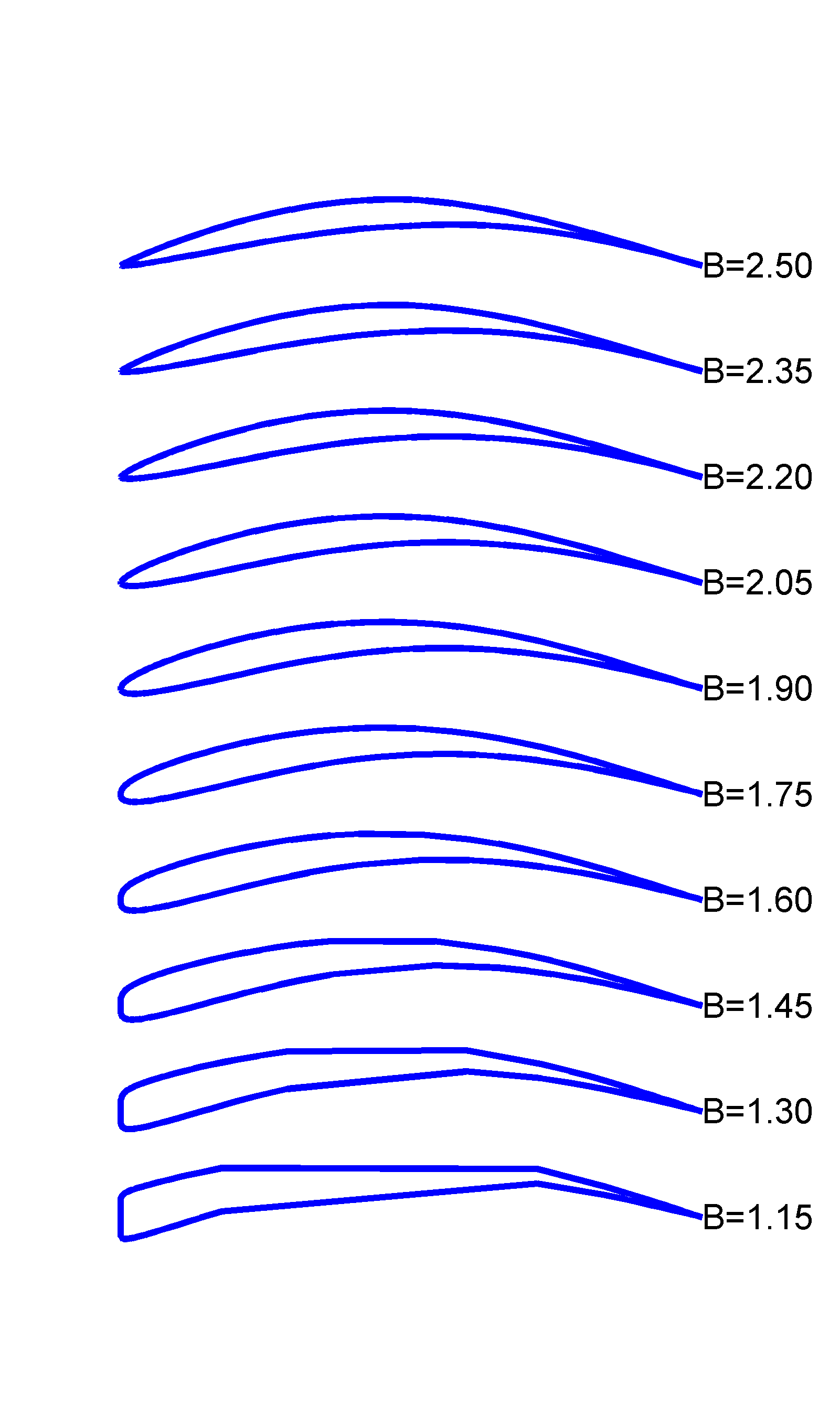}
\includegraphics[width=.15\linewidth]{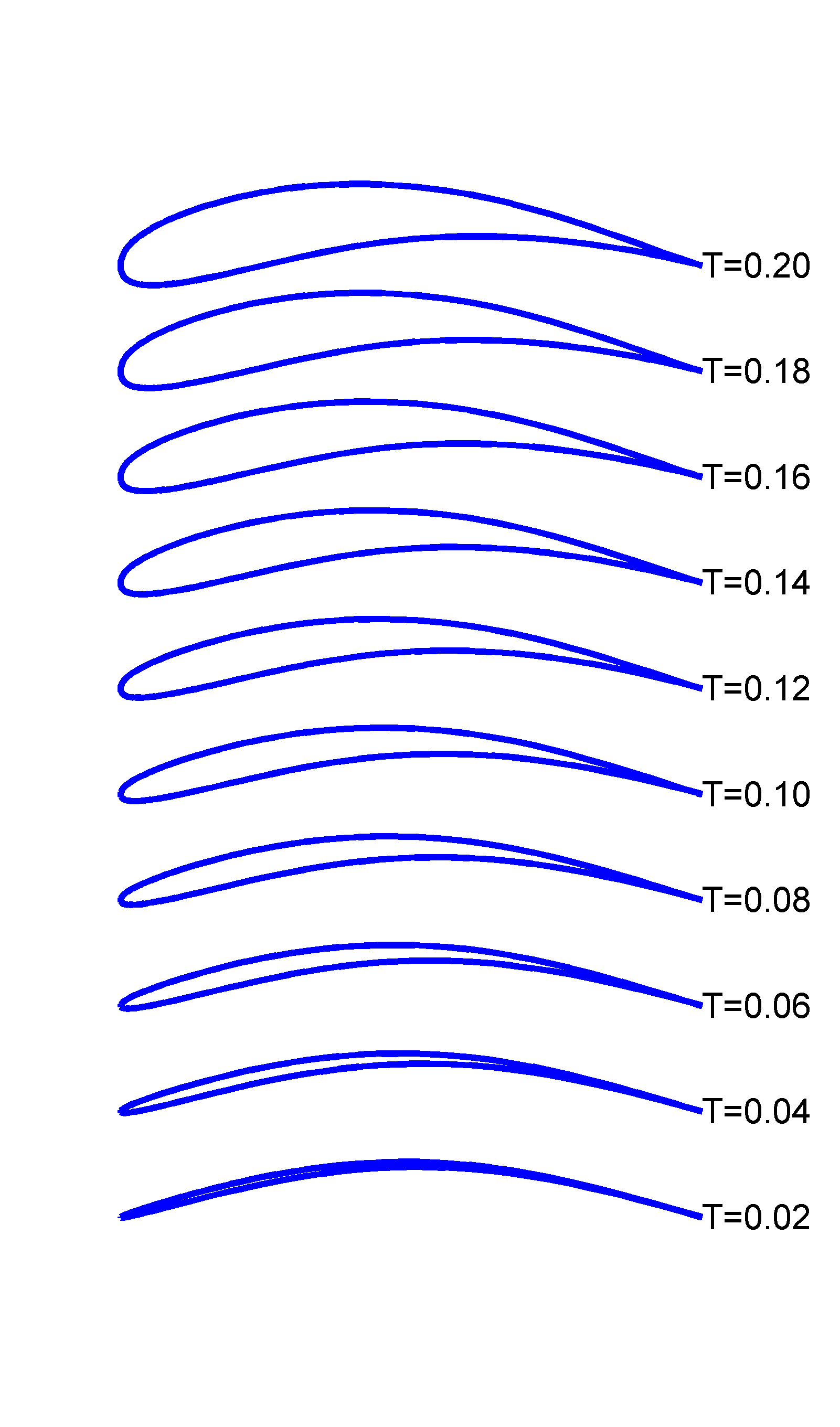}
\includegraphics[width=.15\linewidth]{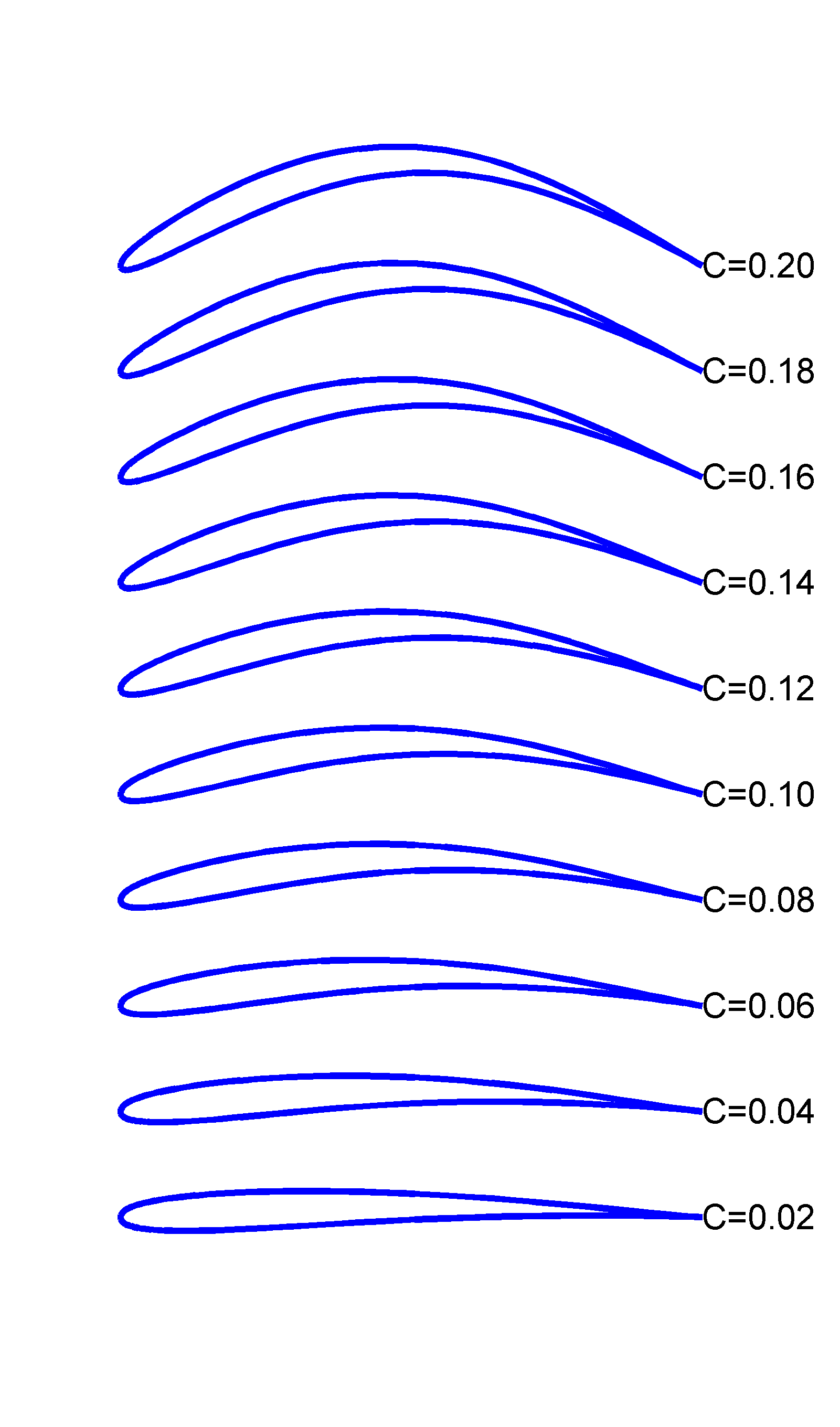}
\includegraphics[width=.15\linewidth]{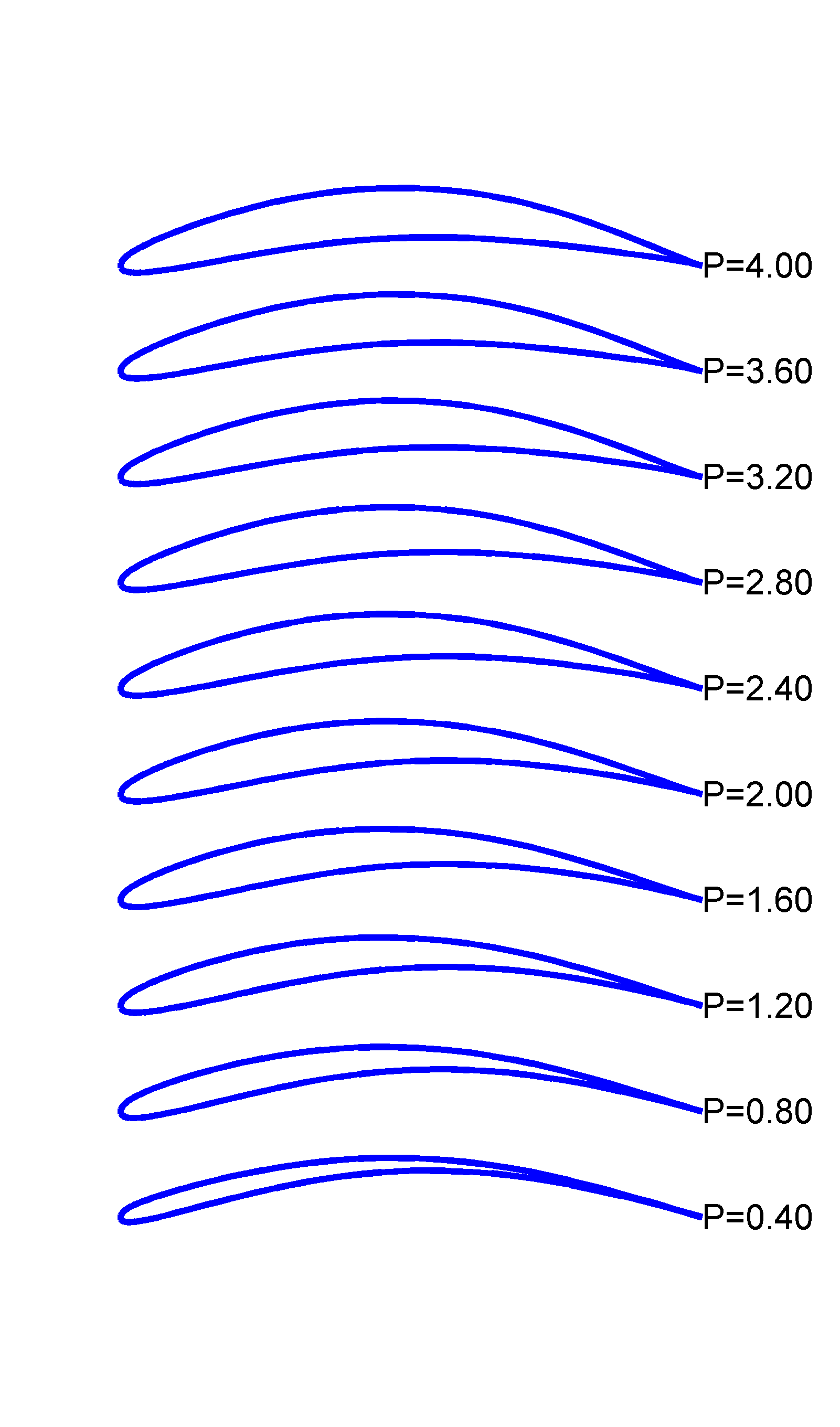}
\includegraphics[width=.15\linewidth]{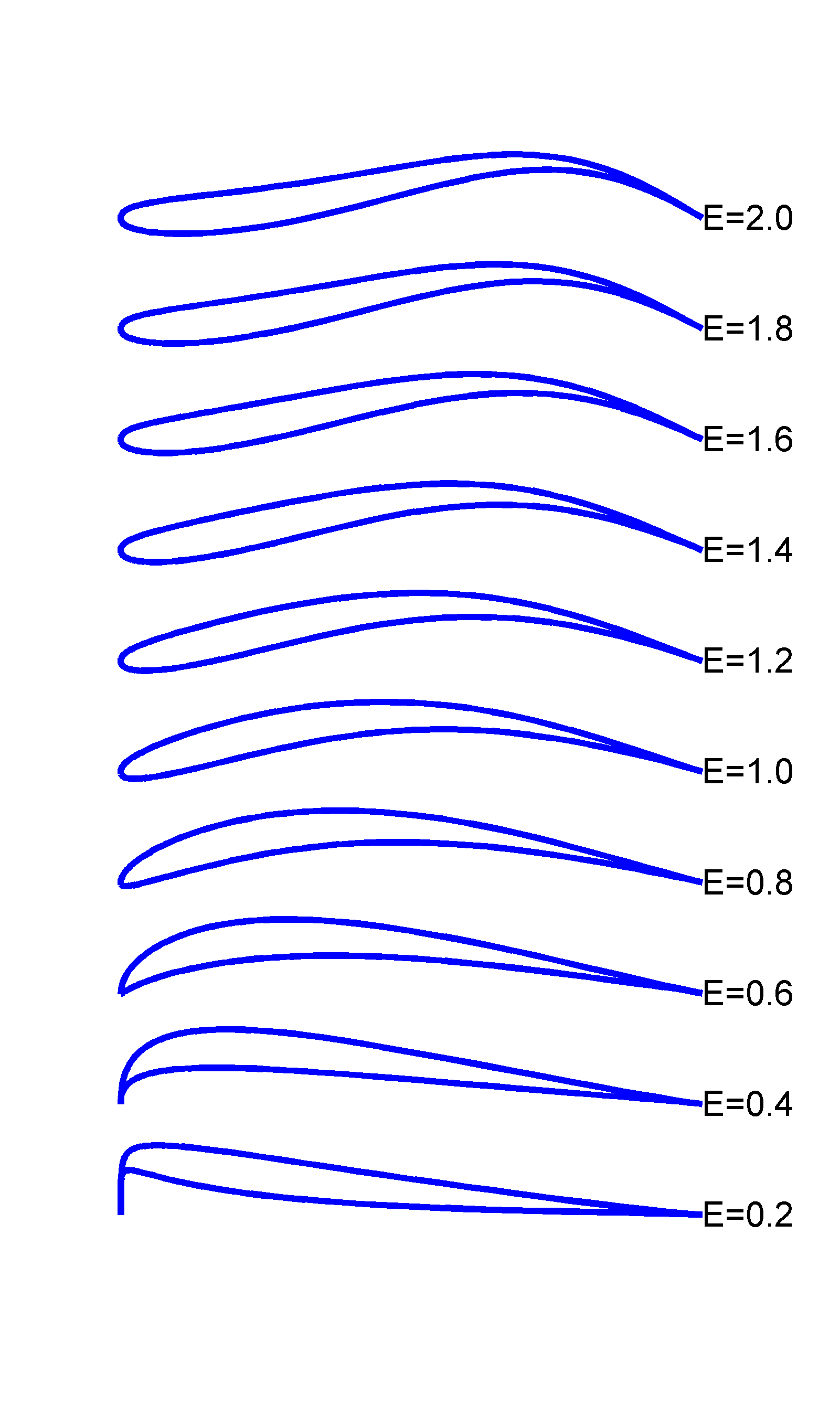}
\caption{Airfoil shapes generated for $B=2$, $T=0.1$, $C=0.05$, $P=1$, $E=1$, $R=0$, where one of the parameters is changed according to the description near the profile.}\label{Fig_foils}
\end{figure}

It should be stressed that the presented description allows for intuitive manipulation of airfoil shape. The parameters have a clear meaning, and their change impacts the profile in a predictable manner. Moreover, the generated shape is continuous and relatively smooth. The example airfoils generated with the above equation are shown on the Fig. \ref{Fig_foils}. One can make several general observations. All the parameters affect the airfoil globally. One of the crucial characteristics of the airfoil is the leading edge, which is strongly dependent on the value of $B$, with a typical, rounded shape obtained for $B \approx 2$. The taper parameter $P$ provides control over the maximum thickness point; a characteristic shape of low drag laminar flow airfoils\cite{Selig} can be obtained for high values of $P$. The thickness distribution $Th$ is a product of $T$ and $P$, and it is given by
\begin{equation}
Th(\theta)=T\frac{|\sin \theta|^{B}}{\sin \theta}(1-X^P),
\end{equation}
for $\theta=\langle 0,\pi \rangle$ and $X$ given by Eq. \ref{Main_eq}. Likewise, one can also define the camber line $Cm$
\begin{equation}
Cm(\theta)=C\sin(X^E\pi)+R\sin(X2\pi).
\end{equation}
These two parameters give an insight on the design performance characteristic and allow for easy comparison with other airfoils.
\section{Optimization procedure}
\begin{figure}[ht!]
\includegraphics[width=.65\linewidth]{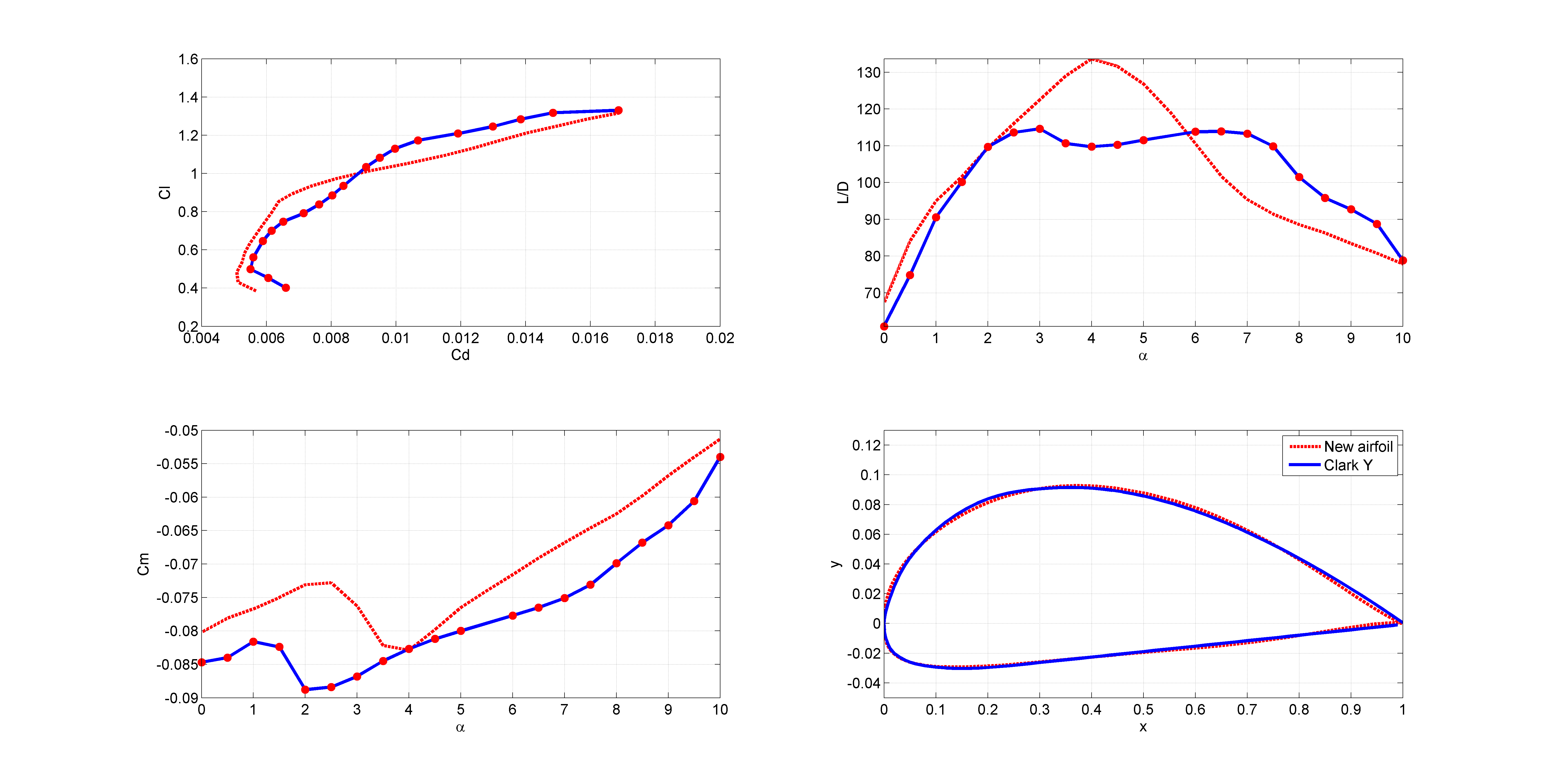}
\caption{The result of shape fitting to the Clark Y airfoil, showing the profile and calculated performance comparison.}\label{Fig_clark}
\end{figure}

\begin{figure}[ht!]
\includegraphics[width=.65\linewidth]{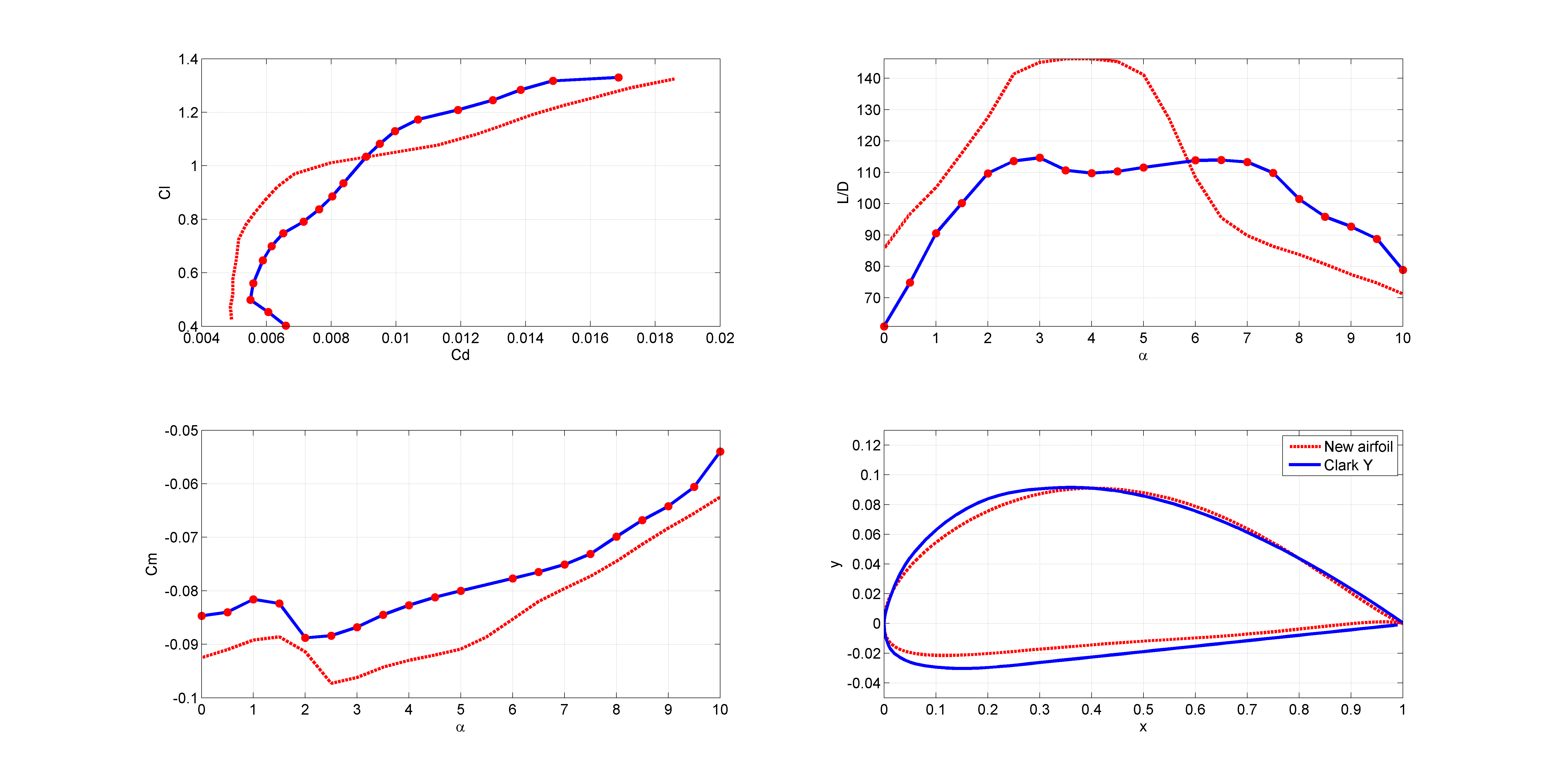}
\caption{The result of shape evolution based on the Clark Y airfoil, showing the profile and calculated performance comparison. The parameters are $B=1.8761$, $T=0.1138$, $P=3,041$, $C=0.03869$, $E=0.8510$, $R=0$.}\label{Fig_clark2}
\end{figure}

The performance optimization is performed in several steps. A known airfoil is chosen as a reference. First, the shape parameters are adjusted to match the selected profile. Due to the limited precision of the curve fitting and the constraints resulting from the relatively low number of free parameters of Eq. \ref{Fig_foils}, small differences between generated shape and reference one are expected. This is shown on the Fig. \ref{Fig_clark}, bottom right panel, where a commonly used Clark Y profile \cite{Clark} has been selected as the base shape. As a next step, the analysis tool XFOIL by Mark Drela \cite{Drela} is used to evaluate the lift and drag polars of the resulting shape and compare it with the chosen airfoil at a given operating point. The results for Reynolds number $Re=10^6$ are shown on the Fig. \ref{Fig_clark}. Due to the highly nonlinear nature of the problem, relatively small deviations from the reference shape result in significant differences in the performance figures. The lift-drag polar is shown on the upper left panel. The obtained airfoil maintains general characteristics of Clark Y - the minimum coefficient of drag $Cd$ is obtained at lift coefficient $Cl \approx 0.5$ and starts to increase significantly for $Cl \approx 0.8$. Interestingly, the generated profile is characterized by notably smaller drag at $Cl \approx 0.9$, which contributes to higher maximum lift to drag ratio (marked as $L/D$, upper right panel, the peak is reached for angle of attack $\alpha = 4^o$). Moreover, the new airfoil has the advantage of a lower pitching moment $Cm$ (bottom left panel). As a final step, the optimization of parameters is performed, with the aim of maximization of $L/D$ over a range of angles of attack. The evolutionary algorithm is used to alter the shape, using the obtained profile as a seed upon which new airfoils are generated. The results of optimization are shown on the Fig. \ref{Fig_clark2}. The peak $L/D$ is significantly increased and high values of lift to drag ratio are obtained at a bigger range of angles of attack. One can see that the evolutionary optimization allowed for drag reduction, while keeping maximum lift coefficient and pitching moment relatively constant. In the optimization, the genetic algorithm ran for 10 generations, each consisting of 9 airfoils. Notably, due to the smooth nature of the generated shapes, there were no cases of a very bad performance, and variation of peak $L/D$ among the population was not very significant. 

\begin{figure}[ht!]
\includegraphics[width=.7\linewidth]{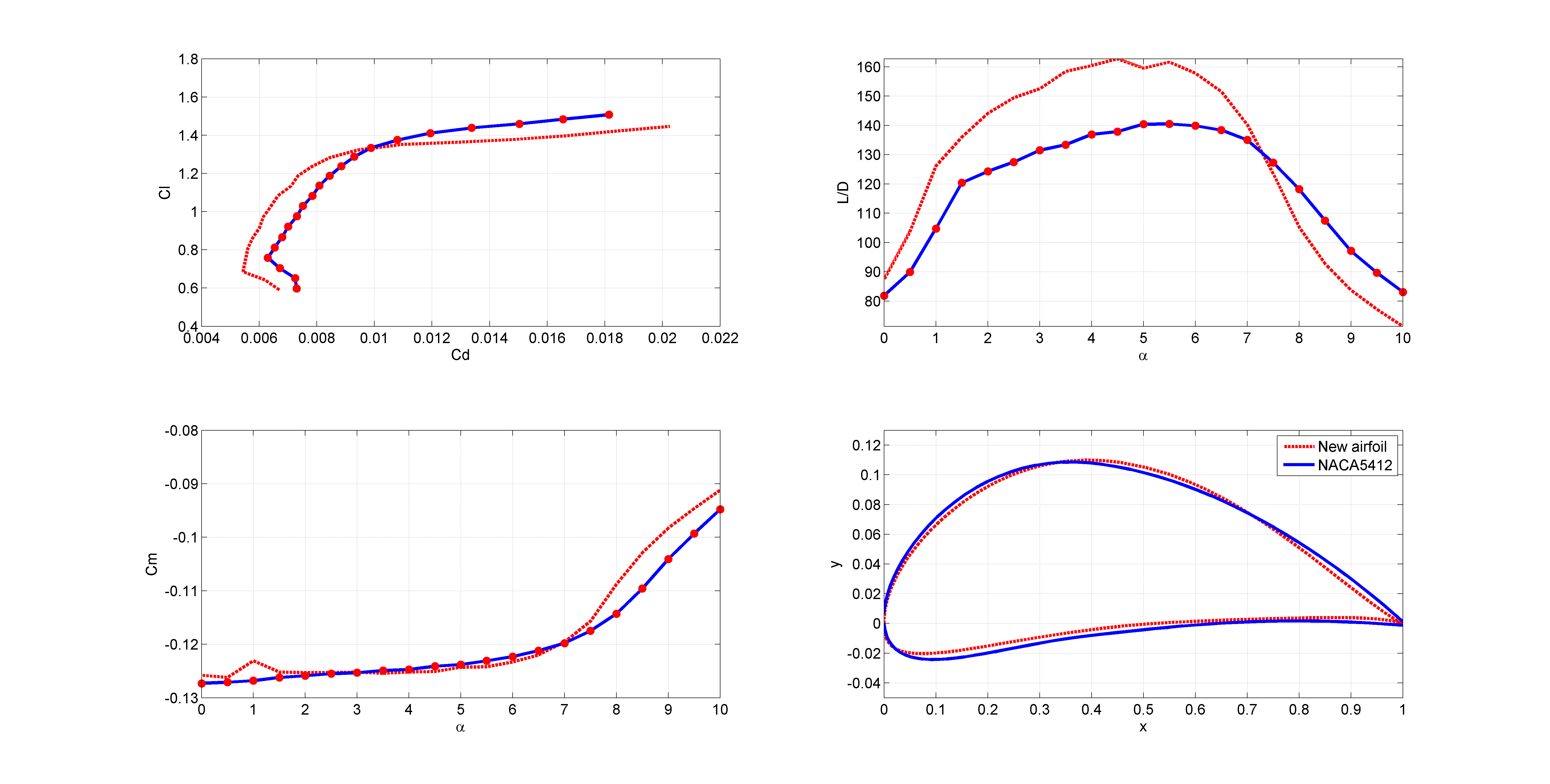}
\caption{The result of shape evolution based on the NACA5412 airfoil, showing the profile and calculated performance comparison. The parameters are $B=1.8608$, $T=0.1277$, $P=2.5536$, $C=0.05332$, $E=0.8434$, $R=0$.}\label{Fig_naca1}
\end{figure}

One can conclude that the procedure is capable of improving upon existing airfoil while keeping its key characteristics, provided that the initial shape can be closely matched by Eq. \ref{Main_eq}. As another example, where clear performance improvement has been obtained, we have used NACA 4 digit profile \cite{naca}, the NACA5412. The optimization results are shown on the Fig. \ref{Fig_naca1}. A significant increase of $L/D$ over the whole angle of attack range is gained with relatively small alteration to the initial shape, while having almost no impact on pitching moment curve. It should be stressed that calculations were stopped before local optimum has been found and further refinement is still possible. However, longer evolution would divert more from the starting shape and would possibly need additional constraints on the geometry of generated airfoil to keep the assumed design criteria.

\begin{figure}[ht!]
\includegraphics[width=.7\linewidth]{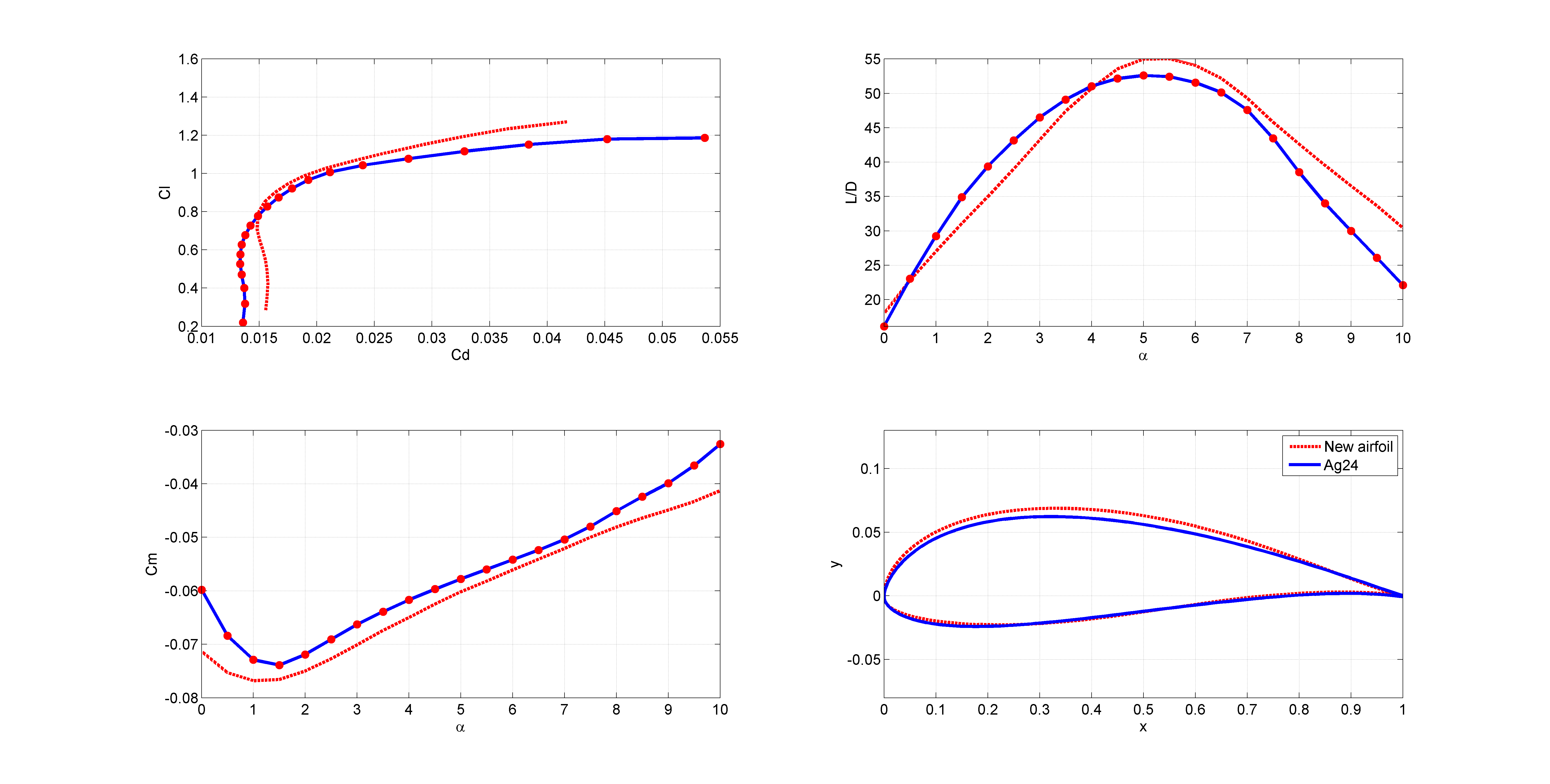}
\caption{The result of shape evolution based on the AG24 airfoil by Mark Drela. The parameters are $B=1.9731$, $T=0.1176$, $P=1.4890$, $C=0.0277$, $E=0.6553$, $R=-0.0042$.}\label{Fig_drela}
\end{figure}

An example of overspecialisation due to the lack of sufficient constraints is shown the Fig. \ref{Fig_drela}. The evolution of low Reynolds number airfoil AG24 by Mark Drela \cite{Summary} results in a design compromise which sacrifices low drag at small lift coefficient for a small gain in peak $L/D$.

\begin{figure}[ht!]
\includegraphics[width=.7\linewidth]{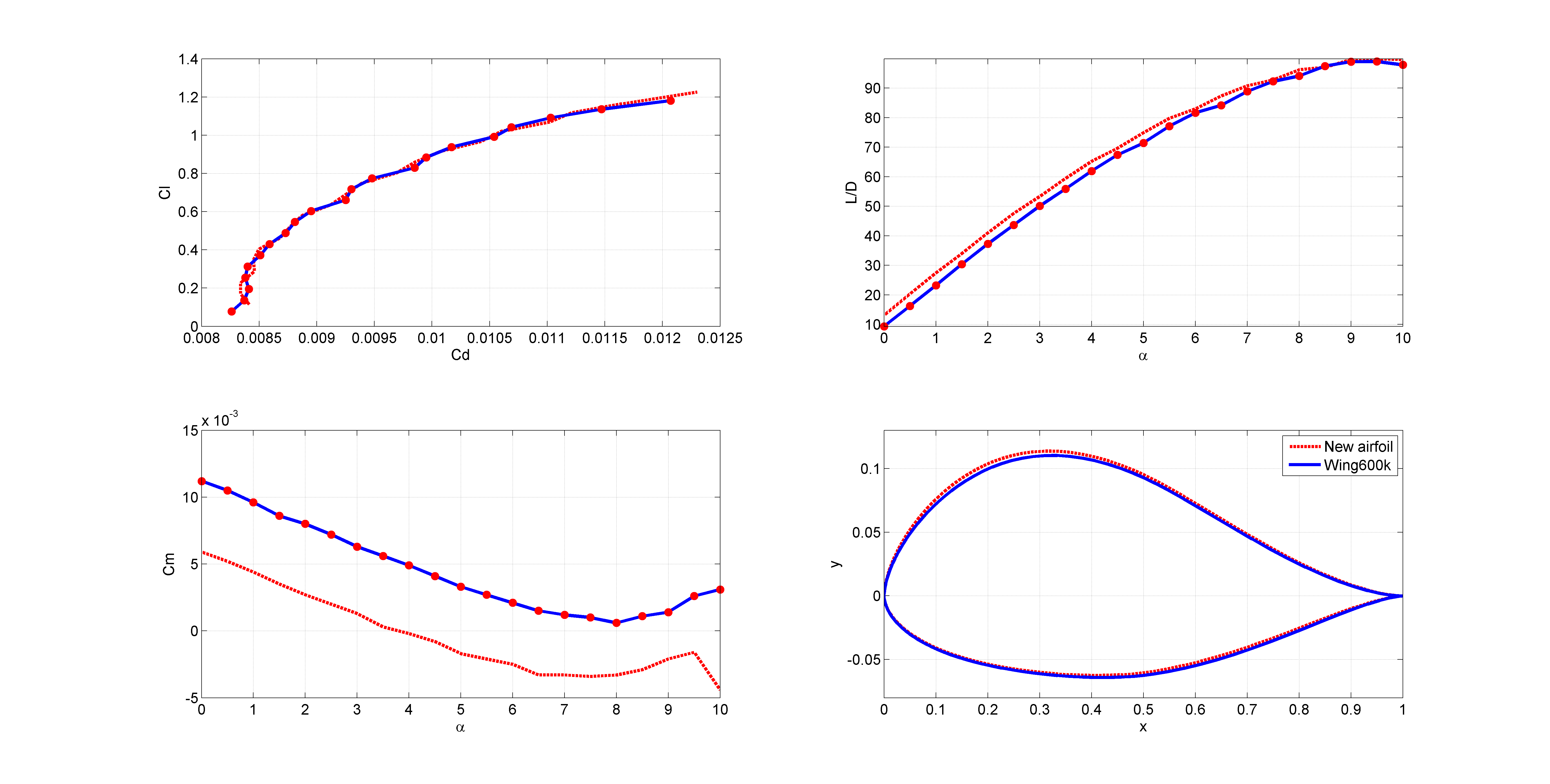}
\caption{Airfoil design for a flyng wing. The parameters are $B=2.1548$, $T=0.2309$, $P=1.6202$, $C=0.0194$, $E=0.6304$, $R=0.0078$.}\label{Fig_wing}
\end{figure}

Finally, the Fig. \ref{Fig_wing} shows the results for a brand new airfoil created for a flying wing. In this example, the baseline shape has been generated by manual adjustment of the parameters according to the design goals of $18\%$ thickness and near zero pitching moment at $Re=6\cdot 10^5$. The optimization has been used to perform only fine adjustments, further refining the design.

\clearpage
\section{Conclusions}
We have created a simple mathematical model for description of the airfoil shape with only $6$ parameters, which allows the designer to easily numerically describe the curve by using a simple terms like thickness, camber, reflex. The description is flexible enough to approximate a wide range of existing profiles, while its simplicity is well suited for optimization with a genetic algorithm. For a wide range of parameters, the obtained shape is smooth and lacks discontinuities, making it attractive for easy numerical analysis. We show the procedure that can be used to improve upon existing, well known airfoils at specific operating points by creating a close approximation with provided model and then refining it. We also show an example of another hybrid approach, where a new design tailored specifically for a flying wing type aircraft is designed manually and further improved by genetic algorithm.

\clearpage
\appendix
\section{Airfoil coordinates}
\begin{table}[ht!]
\caption{Airfoil coordinates for designs based on NACA5412, Clark Y, Ag 24, and flying wing profile.}
\label{my-label}
\resizebox{0.095\linewidth}{!}{
\begin{minipage}{.25\linewidth}
\begin{tabular}{|l|l|}
\hline
X & Y \\ \hline
1.00000 & 0.00000  \\ \hline
0.99770 & 0.00038  \\ \hline
0.99104 & 0.00163  \\ \hline
0.98039 & 0.00388  \\ \hline
0.96610 & 0.00718  \\ \hline
0.94855 & 0.01150  \\ \hline
0.92809 & 0.01678  \\ \hline
0.90504 & 0.02292  \\ \hline
0.87973 & 0.02977  \\ \hline
0.85246 & 0.03716  \\ \hline
0.82348 & 0.04492  \\ \hline
0.79306 & 0.05287  \\ \hline
0.76138 & 0.06082  \\ \hline
0.72862 & 0.06861  \\ \hline
0.69491 & 0.07608  \\ \hline
0.66029 & 0.08310  \\ \hline
0.62471 & 0.08956  \\ \hline
0.58791 & 0.09537  \\ \hline
0.54911 & 0.10048  \\ \hline
0.50412 & 0.10508  \\ \hline
0.45680 & 0.10838  \\ \hline
0.42049 & 0.10975  \\ \hline
0.38793 & 0.11004  \\ \hline
0.35802 & 0.10949  \\ \hline
0.33028 & 0.10825  \\ \hline
0.30444 & 0.10645  \\ \hline
0.28031 & 0.10417  \\ \hline
0.25775 & 0.10152  \\ \hline
0.23666 & 0.09856  \\ \hline
0.21695 & 0.09535  \\ \hline
0.19853 & 0.09197  \\ \hline
0.18134 & 0.08845  \\ \hline
0.16530 & 0.08484  \\ \hline
0.15036 & 0.08117  \\ \hline
0.13645 & 0.07748  \\ \hline
0.12351 & 0.07379  \\ \hline
0.11150 & 0.07011  \\ \hline
0.10035 & 0.06647  \\ \hline
0.09003 & 0.06288  \\ \hline
0.08049 & 0.05935  \\ \hline
0.07167 & 0.05588  \\ \hline
0.06354 & 0.05249  \\ \hline
0.05607 & 0.04917  \\ \hline
0.04920 & 0.04593  \\ \hline
0.04290 & 0.04277  \\ \hline
0.03715 & 0.03969  \\ \hline
0.03191 & 0.03669  \\ \hline
0.02716 & 0.03376  \\ \hline
0.02285 & 0.03090  \\ \hline
0.01899 & 0.02812  \\ \hline
0.01553 & 0.02540  \\ \hline
0.01246 & 0.02274  \\ \hline
0.00976 & 0.02014  \\ \hline
0.00741 & 0.01759  \\ \hline
0.00541 & 0.01509  \\ \hline
0.00374 & 0.01263  \\ \hline
0.00238 & 0.01020  \\ \hline
0.00133 & 0.00778  \\ \hline
0.00059 & 0.00536  \\ \hline
0.00015 & 0.00287  \\ \hline
0.00000 & 0.00000  \\ \hline
0.00015 & -0.00268 \\ \hline
0.00059 & -0.00473 \\ \hline
0.00133 & -0.00652 \\ \hline
0.00238 & -0.00815 \\ \hline
0.00374 & -0.00963 \\ \hline
0.00541 & -0.01099 \\ \hline
0.00741 & -0.01224 \\ \hline
0.00976 & -0.01340 \\ \hline
0.01246 & -0.01445 \\ \hline
0.01553 & -0.01542 \\ \hline
0.01899 & -0.01631 \\ \hline
0.02285 & -0.01710 \\ \hline
0.02716 & -0.01782 \\ \hline
0.03191 & -0.01844 \\ \hline
0.03715 & -0.01898 \\ \hline
0.04290 & -0.01943 \\ \hline
0.04920 & -0.01979 \\ \hline
0.05607 & -0.02005 \\ \hline
0.06354 & -0.02022 \\ \hline
0.07167 & -0.02030 \\ \hline
0.08049 & -0.02027 \\ \hline
0.09003 & -0.02014 \\ \hline
0.10035 & -0.01990 \\ \hline
0.11150 & -0.01956 \\ \hline
0.12351 & -0.01911 \\ \hline
0.13645 & -0.01854 \\ \hline
0.15036 & -0.01787 \\ \hline
0.16530 & -0.01709 \\ \hline
0.18134 & -0.01621 \\ \hline
0.19853 & -0.01522 \\ \hline
0.21695 & -0.01413 \\ \hline
0.23666 & -0.01296 \\ \hline
0.25775 & -0.01171 \\ \hline
0.28031 & -0.01039 \\ \hline
0.30444 & -0.00901 \\ \hline
0.33028 & -0.00760 \\ \hline
0.35802 & -0.00616 \\ \hline
0.38793 & -0.00472 \\ \hline
0.42049 & -0.00329 \\ \hline
0.45680 & -0.00189 \\ \hline
0.50412 & -0.00041 \\ \hline
0.54911 & 0.00061  \\ \hline
0.58791 & 0.00128  \\ \hline
0.62471 & 0.00181  \\ \hline
0.66029 & 0.00225  \\ \hline
0.69491 & 0.00264  \\ \hline
0.72862 & 0.00300  \\ \hline
0.76138 & 0.00332  \\ \hline
0.79306 & 0.00360  \\ \hline
0.82348 & 0.00382  \\ \hline
0.85246 & 0.00395  \\ \hline
0.87973 & 0.00396  \\ \hline
0.90504 & 0.00383  \\ \hline
0.92809 & 0.00353  \\ \hline
0.94855 & 0.00305  \\ \hline
0.96610 & 0.00241  \\ \hline
0.98039 & 0.00166  \\ \hline
0.99104 & 0.00090  \\ \hline
0.99770 & 0.00027  \\ \hline
1.00000 & 0.00000  \\ \hline
\end{tabular}
\end{minipage}}
\resizebox{0.095\linewidth}{!}{
\begin{minipage}{.25\linewidth}
\begin{tabular}{|l|l|}
\hline
X & Y \\ \hline
1.00000 & 0.00000  \\ \hline
0.99766 & 0.00030  \\ \hline
0.99088 & 0.00133  \\ \hline
0.98004 & 0.00324  \\ \hline
0.96552 & 0.00608  \\ \hline
0.94769 & 0.00986  \\ \hline
0.92691 & 0.01450  \\ \hline
0.90353 & 0.01991  \\ \hline
0.87788 & 0.02592  \\ \hline
0.85027 & 0.03238  \\ \hline
0.82099 & 0.03911  \\ \hline
0.79029 & 0.04593  \\ \hline
0.75838 & 0.05268  \\ \hline
0.72547 & 0.05921  \\ \hline
0.69167 & 0.06537  \\ \hline
0.65708 & 0.07106  \\ \hline
0.62167 & 0.07620  \\ \hline
0.58524 & 0.08071  \\ \hline
0.54713 & 0.08456  \\ \hline
0.50378 & 0.08788  \\ \hline
0.45864 & 0.09015  \\ \hline
0.42304 & 0.09104  \\ \hline
0.39087 & 0.09110  \\ \hline
0.36116 & 0.09052  \\ \hline
0.33351 & 0.08940  \\ \hline
0.30768 & 0.08784  \\ \hline
0.28350 & 0.08592  \\ \hline
0.26085 & 0.08371  \\ \hline
0.23965 & 0.08126  \\ \hline
0.21980 & 0.07863  \\ \hline
0.20123 & 0.07586  \\ \hline
0.18388 & 0.07299  \\ \hline
0.16768 & 0.07005  \\ \hline
0.15257 & 0.06707  \\ \hline
0.13850 & 0.06407  \\ \hline
0.12540 & 0.06107  \\ \hline
0.11324 & 0.05808  \\ \hline
0.10194 & 0.05512  \\ \hline
0.09148 & 0.05220  \\ \hline
0.08179 & 0.04932  \\ \hline
0.07285 & 0.04650  \\ \hline
0.06460 & 0.04373  \\ \hline
0.05700 & 0.04101  \\ \hline
0.05003 & 0.03836  \\ \hline
0.04363 & 0.03576  \\ \hline
0.03779 & 0.03322  \\ \hline
0.03246 & 0.03075  \\ \hline
0.02762 & 0.02833  \\ \hline
0.02325 & 0.02596  \\ \hline
0.01932 & 0.02365  \\ \hline
0.01580 & 0.02138  \\ \hline
0.01267 & 0.01916  \\ \hline
0.00993 & 0.01699  \\ \hline
0.00754 & 0.01485  \\ \hline
0.00551 & 0.01275  \\ \hline
0.00380 & 0.01068  \\ \hline
0.00242 & 0.00862  \\ \hline
0.00136 & 0.00657  \\ \hline
0.00060 & 0.00452  \\ \hline
0.00015 & 0.00241  \\ \hline
0.00000 & 0.00000  \\ \hline
0.00015 & -0.00227 \\ \hline
0.00060 & -0.00408 \\ \hline
0.00136 & -0.00569 \\ \hline
0.00242 & -0.00718 \\ \hline
0.00380 & -0.00856 \\ \hline
0.00551 & -0.00985 \\ \hline
0.00754 & -0.01106 \\ \hline
0.00993 & -0.01219 \\ \hline
0.01267 & -0.01326 \\ \hline
0.01580 & -0.01427 \\ \hline
0.01932 & -0.01521 \\ \hline
0.02325 & -0.01609 \\ \hline
0.02762 & -0.01690 \\ \hline
0.03246 & -0.01766 \\ \hline
0.03779 & -0.01835 \\ \hline
0.04363 & -0.01898 \\ \hline
0.05003 & -0.01954 \\ \hline
0.05700 & -0.02004 \\ \hline
0.06460 & -0.02047 \\ \hline
0.07285 & -0.02083 \\ \hline
0.08179 & -0.02111 \\ \hline
0.09148 & -0.02133 \\ \hline
0.10194 & -0.02146 \\ \hline
0.11324 & -0.02152 \\ \hline
0.12540 & -0.02150 \\ \hline
0.13850 & -0.02140 \\ \hline
0.15257 & -0.02122 \\ \hline
0.16768 & -0.02096 \\ \hline
0.18388 & -0.02062 \\ \hline
0.20123 & -0.02021 \\ \hline
0.21980 & -0.01972 \\ \hline
0.23965 & -0.01917 \\ \hline
0.26085 & -0.01855 \\ \hline
0.28350 & -0.01787 \\ \hline
0.30768 & -0.01714 \\ \hline
0.33351 & -0.01637 \\ \hline
0.36116 & -0.01555 \\ \hline
0.39087 & -0.01469 \\ \hline
0.42304 & -0.01380 \\ \hline
0.45864 & -0.01286 \\ \hline
0.50378 & -0.01178 \\ \hline
0.54713 & -0.01086 \\ \hline
0.58524 & -0.01007 \\ \hline
0.62167 & -0.00926 \\ \hline
0.65708 & -0.00839 \\ \hline
0.69167 & -0.00745 \\ \hline
0.72547 & -0.00642 \\ \hline
0.75838 & -0.00531 \\ \hline
0.79029 & -0.00416 \\ \hline
0.82099 & -0.00300 \\ \hline
0.85027 & -0.00189 \\ \hline
0.87788 & -0.00087 \\ \hline
0.90353 & -0.00002 \\ \hline
0.92691 & 0.00060  \\ \hline
0.94769 & 0.00097  \\ \hline
0.96552 & 0.00106  \\ \hline
0.98004 & 0.00090  \\ \hline
0.99088 & 0.00056  \\ \hline
0.99766 & 0.00019  \\ \hline
1.00000 & 0.00000  \\ \hline
\end{tabular}
\end{minipage}}
\resizebox{0.095\linewidth}{!}{
\begin{minipage}{.25\linewidth}
\begin{tabular}{|l|l|}
\hline
X & Y \\ \hline
1.00000 & 0.00000  \\ \hline
0.99740 & 0.00024  \\ \hline
0.98988 & 0.00103  \\ \hline
0.97788 & 0.00244  \\ \hline
0.96185 & 0.00450  \\ \hline
0.94224 & 0.00721  \\ \hline
0.91950 & 0.01053  \\ \hline
0.89406 & 0.01439  \\ \hline
0.86634 & 0.01868  \\ \hline
0.83674 & 0.02330  \\ \hline
0.80562 & 0.02810  \\ \hline
0.77333 & 0.03296  \\ \hline
0.74017 & 0.03775  \\ \hline
0.70644 & 0.04237  \\ \hline
0.67237 & 0.04673  \\ \hline
0.63818 & 0.05076  \\ \hline
0.60405 & 0.05441  \\ \hline
0.57007 & 0.05765  \\ \hline
0.53628 & 0.06049  \\ \hline
0.50220 & 0.06295  \\ \hline
0.46862 & 0.06500  \\ \hline
0.43745 & 0.06653  \\ \hline
0.40779 & 0.06766  \\ \hline
0.37952 & 0.06841  \\ \hline
0.35260 & 0.06881  \\ \hline
0.32698 & 0.06890  \\ \hline
0.30266 & 0.06869  \\ \hline
0.27960 & 0.06821  \\ \hline
0.25779 & 0.06747  \\ \hline
0.23720 & 0.06649  \\ \hline
0.21779 & 0.06530  \\ \hline
0.19953 & 0.06392  \\ \hline
0.18238 & 0.06236  \\ \hline
0.16630 & 0.06065  \\ \hline
0.15125 & 0.05880  \\ \hline
0.13719 & 0.05682  \\ \hline
0.12408 & 0.05475  \\ \hline
0.11187 & 0.05258  \\ \hline
0.10052 & 0.05035  \\ \hline
0.08998 & 0.04805  \\ \hline
0.08023 & 0.04571  \\ \hline
0.07122 & 0.04334  \\ \hline
0.06290 & 0.04094  \\ \hline
0.05525 & 0.03853  \\ \hline
0.04822 & 0.03611  \\ \hline
0.04179 & 0.03370  \\ \hline
0.03592 & 0.03129  \\ \hline
0.03059 & 0.02889  \\ \hline
0.02576 & 0.02650  \\ \hline
0.02141 & 0.02414  \\ \hline
0.01752 & 0.02179  \\ \hline
0.01406 & 0.01947  \\ \hline
0.01102 & 0.01718  \\ \hline
0.00837 & 0.01491  \\ \hline
0.00611 & 0.01267  \\ \hline
0.00422 & 0.01046  \\ \hline
0.00269 & 0.00829  \\ \hline
0.00151 & 0.00614  \\ \hline
0.00067 & 0.00404  \\ \hline
0.00017 & 0.00198  \\ \hline
0.00000 & 0.00000  \\ \hline
0.00017 & -0.00141 \\ \hline
0.00067 & -0.00263 \\ \hline
0.00151 & -0.00376 \\ \hline
0.00269 & -0.00483 \\ \hline
0.00422 & -0.00586 \\ \hline
0.00611 & -0.00684 \\ \hline
0.00837 & -0.00780 \\ \hline
0.01102 & -0.00873 \\ \hline
0.01406 & -0.00964 \\ \hline
0.01752 & -0.01053 \\ \hline
0.02141 & -0.01139 \\ \hline
0.02576 & -0.01225 \\ \hline
0.03059 & -0.01308 \\ \hline
0.03592 & -0.01390 \\ \hline
0.04179 & -0.01470 \\ \hline
0.04822 & -0.01548 \\ \hline
0.05525 & -0.01625 \\ \hline
0.06290 & -0.01700 \\ \hline
0.07122 & -0.01772 \\ \hline
0.08023 & -0.01842 \\ \hline
0.08998 & -0.01910 \\ \hline
0.10052 & -0.01974 \\ \hline
0.11187 & -0.02034 \\ \hline
0.12408 & -0.02090 \\ \hline
0.13719 & -0.02140 \\ \hline
0.15125 & -0.02185 \\ \hline
0.16630 & -0.02223 \\ \hline
0.18238 & -0.02252 \\ \hline
0.19953 & -0.02272 \\ \hline
0.21779 & -0.02281 \\ \hline
0.23720 & -0.02277 \\ \hline
0.25779 & -0.02259 \\ \hline
0.27960 & -0.02225 \\ \hline
0.30266 & -0.02174 \\ \hline
0.32698 & -0.02104 \\ \hline
0.35260 & -0.02014 \\ \hline
0.37952 & -0.01902 \\ \hline
0.40779 & -0.01770 \\ \hline
0.43745 & -0.01616 \\ \hline
0.46862 & -0.01443 \\ \hline
0.50220 & -0.01247 \\ \hline
0.53628 & -0.01045 \\ \hline
0.57007 & -0.00844 \\ \hline
0.60405 & -0.00647 \\ \hline
0.63818 & -0.00457 \\ \hline
0.67237 & -0.00280 \\ \hline
0.70644 & -0.00121 \\ \hline
0.74017 & 0.00016  \\ \hline
0.77333 & 0.00128  \\ \hline
0.80562 & 0.00212  \\ \hline
0.83674 & 0.00269  \\ \hline
0.86634 & 0.00298  \\ \hline
0.89406 & 0.00301  \\ \hline
0.91950 & 0.00281  \\ \hline
0.94224 & 0.00241  \\ \hline
0.96185 & 0.00187  \\ \hline
0.97788 & 0.00125  \\ \hline
0.98988 & 0.00066  \\ \hline
0.99740 & 0.00019  \\ \hline
1.00000 & 0.00000  \\ \hline
\end{tabular}
\end{minipage}}
\resizebox{0.095\linewidth}{!}{
\begin{minipage}{.25\linewidth}
\begin{tabular}{|l|l|}
\hline
X & Y \\ \hline
1.00000 & 0.00000  \\ \hline
0.99692 & 0.00001  \\ \hline
0.98802 & 0.00023  \\ \hline
0.97386 & 0.00093  \\ \hline
0.95506 & 0.00234  \\ \hline
0.93222 & 0.00462  \\ \hline
0.90598 & 0.00788  \\ \hline
0.87693 & 0.01213  \\ \hline
0.84567 & 0.01738  \\ \hline
0.81278 & 0.02356  \\ \hline
0.77878 & 0.03053  \\ \hline
0.74418 & 0.03815  \\ \hline
0.70944 & 0.04621  \\ \hline
0.67501 & 0.05448  \\ \hline
0.64129 & 0.06273  \\ \hline
0.60868 & 0.07072  \\ \hline
0.57761 & 0.07820  \\ \hline
0.54855 & 0.08495  \\ \hline
0.52223 & 0.09075  \\ \hline
0.50080 & 0.09514  \\ \hline
0.48128 & 0.09869  \\ \hline
0.45757 & 0.10254  \\ \hline
0.43275 & 0.10602  \\ \hline
0.40764 & 0.10894  \\ \hline
0.38266 & 0.11120  \\ \hline
0.35808 & 0.11275  \\ \hline
0.33411 & 0.11357  \\ \hline
0.31086 & 0.11369  \\ \hline
0.28845 & 0.11312  \\ \hline
0.26694 & 0.11192  \\ \hline
0.24637 & 0.11013  \\ \hline
0.22678 & 0.10783  \\ \hline
0.20818 & 0.10507  \\ \hline
0.19057 & 0.10192  \\ \hline
0.17394 & 0.09843  \\ \hline
0.15828 & 0.09467  \\ \hline
0.14357 & 0.09070  \\ \hline
0.12979 & 0.08655  \\ \hline
0.11691 & 0.08227  \\ \hline
0.10490 & 0.07791  \\ \hline
0.09372 & 0.07350  \\ \hline
0.08334 & 0.06906  \\ \hline
0.07374 & 0.06463  \\ \hline
0.06487 & 0.06022  \\ \hline
0.05670 & 0.05585  \\ \hline
0.04920 & 0.05154  \\ \hline
0.04234 & 0.04730  \\ \hline
0.03609 & 0.04313  \\ \hline
0.03042 & 0.03905  \\ \hline
0.02530 & 0.03507  \\ \hline
0.02072 & 0.03119  \\ \hline
0.01664 & 0.02742  \\ \hline
0.01305 & 0.02375  \\ \hline
0.00992 & 0.02020  \\ \hline
0.00724 & 0.01678  \\ \hline
0.00501 & 0.01348  \\ \hline
0.00319 & 0.01033  \\ \hline
0.00179 & 0.00734  \\ \hline
0.00079 & 0.00455  \\ \hline
0.00020 & 0.00201  \\ \hline
0.00000 & 0.00000  \\ \hline
0.00020 & -0.00143 \\ \hline
0.00079 & -0.00312 \\ \hline
0.00179 & -0.00491 \\ \hline
0.00319 & -0.00677 \\ \hline
0.00501 & -0.00868 \\ \hline
0.00724 & -0.01063 \\ \hline
0.00992 & -0.01261 \\ \hline
0.01305 & -0.01462 \\ \hline
0.01664 & -0.01666 \\ \hline
0.02072 & -0.01871 \\ \hline
0.02530 & -0.02079 \\ \hline
0.03042 & -0.02288 \\ \hline
0.03609 & -0.02498 \\ \hline
0.04234 & -0.02709 \\ \hline
0.04920 & -0.02920 \\ \hline
0.05670 & -0.03132 \\ \hline
0.06487 & -0.03342 \\ \hline
0.07374 & -0.03552 \\ \hline
0.08334 & -0.03760 \\ \hline
0.09372 & -0.03965 \\ \hline
0.10490 & -0.04168 \\ \hline
0.11691 & -0.04366 \\ \hline
0.12979 & -0.04560 \\ \hline
0.14357 & -0.04749 \\ \hline
0.15828 & -0.04931 \\ \hline
0.17394 & -0.05106 \\ \hline
0.19057 & -0.05273 \\ \hline
0.20818 & -0.05431 \\ \hline
0.22678 & -0.05579 \\ \hline
0.24637 & -0.05716 \\ \hline
0.26694 & -0.05841 \\ \hline
0.28845 & -0.05953 \\ \hline
0.31086 & -0.06050 \\ \hline
0.33411 & -0.06130 \\ \hline
0.35808 & -0.06191 \\ \hline
0.38266 & -0.06231 \\ \hline
0.40764 & -0.06247 \\ \hline
0.43275 & -0.06236 \\ \hline
0.45757 & -0.06196 \\ \hline
0.48128 & -0.06129 \\ \hline
0.50080 & -0.06046 \\ \hline
0.52223 & -0.05916 \\ \hline
0.54855 & -0.05720 \\ \hline
0.57761 & -0.05467 \\ \hline
0.60868 & -0.05156 \\ \hline
0.64129 & -0.04790 \\ \hline
0.67501 & -0.04370 \\ \hline
0.70944 & -0.03903 \\ \hline
0.74418 & -0.03399 \\ \hline
0.77878 & -0.02872 \\ \hline
0.81278 & -0.02339 \\ \hline
0.84567 & -0.01823 \\ \hline
0.87693 & -0.01345 \\ \hline
0.90598 & -0.00925 \\ \hline
0.93222 & -0.00581 \\ \hline
0.95506 & -0.00321 \\ \hline
0.97386 & -0.00146 \\ \hline
0.98802 & -0.00048 \\ \hline
0.99692 & -0.00007 \\ \hline
1.00000 & 0.00000  \\ \hline
\end{tabular}
\end{minipage}}
\end{table}
\end{document}